\documentclass[ALICE,manyauthors]{cernphprep}
\usepackage[comma,square,numbers,sort&compress]{natbib}
\usepackage{hyperref}
\usepackage{lineno}
\usepackage{xspace}
\usepackage{mathptmx}
\usepackage[T1]{fontenc}
\usepackage{orcidlink}

\begin{document}

%

\newcommand{\pp}           {pp\xspace}
\newcommand{\ppbar}        {\mbox{$\mathrm {p\overline{p}}$}\xspace}
\newcommand{\XeXe}         {\mbox{Xe--Xe}\xspace}
\newcommand{\PbPb}         {\mbox{Pb--Pb}\xspace}
\newcommand{\pA}           {\mbox{pA}\xspace}
\newcommand{\pPb}          {\mbox{p--Pb}\xspace}
\newcommand{\AuAu}         {\mbox{Au--Au}\xspace}
\newcommand{\dAu}          {\mbox{d--Au}\xspace}

\newcommand{\s}            {\ensuremath{\sqrt{s}}\xspace}
\newcommand{\snn}          {\ensuremath{\sqrt{s_{\mathrm{NN}}}}\xspace}
\newcommand{\pt}           {\ensuremath{p_{\rm{T}}}\xspace}
\newcommand{\ptsquared}    {\ensuremath{p_{\rm{T}}^{2}}\xspace}
\newcommand{\ptkk}           {\ensuremath{p_{\rm{T, KK}}}\xspace}
\newcommand{\ptsquaredkk}    {\ensuremath{p_{\rm{T, KK}}^{2}}\xspace}
\newcommand{\meanpt}       {$\langle p_{\mathrm{T}}\rangle$\xspace}
\newcommand{\ycms}         {\ensuremath{y_{\rm CMS}}\xspace}
\newcommand{\ylab}         {\ensuremath{y_{\rm lab}}\xspace}
\newcommand{\etarange}[1]  {\mbox{$\left | \eta \right |~<~#1$}}
\newcommand{\yrange}[1]    {\mbox{$\left | y \right |~<~#1$}}
\newcommand{\dndy}         {\ensuremath{\mathrm{d}N_\mathrm{ch}/\mathrm{d}y}\xspace}
\newcommand{\dndeta}       {\ensuremath{\mathrm{d}N_\mathrm{ch}/\mathrm{d}\eta}\xspace}
\newcommand{\avdndeta}     {\ensuremath{\langle\dndeta\rangle}\xspace}
\newcommand{\dNdy}         {\ensuremath{\mathrm{d}N_\mathrm{ch}/\mathrm{d}y}\xspace}
\newcommand{\Npart}        {\ensuremath{N_\mathrm{part}}\xspace}
\newcommand{\Ncoll}        {\ensuremath{N_\mathrm{coll}}\xspace}
\newcommand{\dEdx}         {\ensuremath{\textrm{d}E/\textrm{d}x}\xspace}
\newcommand{\RpPb}         {\ensuremath{R_{\rm pPb}}\xspace}

\newcommand{\nineH}        {$\sqrt{s}~=~0.9$~Te\kern-.1emV\xspace}
\newcommand{\seven}        {$\sqrt{s}~=~7$~Te\kern-.1emV\xspace}
\newcommand{\twoH}         {$\sqrt{s}~=~0.2$~Te\kern-.1emV\xspace}
\newcommand{\twosevensix}  {$\sqrt{s}~=~2.76$~Te\kern-.1emV\xspace}
\newcommand{\five}         {$\sqrt{s}~=~5.02$~Te\kern-.1emV\xspace}
\newcommand{\twosevensixnn}{$\sqrt{s_{\mathrm{NN}}}~=~2.76$~Te\kern-.1emV\xspace}
\newcommand{\fivenn}       {$\sqrt{s_{\mathrm{NN}}}~=~5.02$~Te\kern-.1emV\xspace}
\newcommand{\LT}           {L{\'e}vy-Tsallis\xspace}
\newcommand{\lumi}         {\ensuremath{\mathcal{L}}\xspace}
\newcommand{\GeVc} {Ge\kern-.1emV/$c$\xspace}
\newcommand{\MeVc} {Me\kern-.1emV/$c$\xspace}
\newcommand{\TeV} {Te\kern-.1emV\xspace}
\newcommand{\GeV} {Ge\kern-.1emV\xspace}
\newcommand{\MeV} {Me\kern-.1emV\xspace}
\newcommand{\GeVmass} {Ge\kern-.1emV/$c^2$\xspace}
\newcommand{\MeVmass} {Me\kern-.1emV/$c^2$\xspace}
\newcommand{\GeVmassSR}     {\ensuremath{(\mathrm{Ge\kern-.1emV}/c^{2})^{-1/2}}\ }

\newcommand{\GeVcSquared}   
{\ensuremath{(\mathrm{Ge\kern-.1emV}/c)^{2}}\ }
\newcommand{\inverseGeVcSquared}   {\ensuremath{(\mathrm{Ge\kern-.1emV}/c)^{-2}}\ }
\newcommand{\ITS}          {\rm{ITS}\xspace}
\newcommand{\TOF}          {\rm{TOF}\xspace}
\newcommand{\ZDC}          {\rm{ZDC}\xspace}
\newcommand{\ZDCs}         {\rm{ZDCs}\xspace}
\newcommand{\ZNA}          {\rm{ZNA}\xspace}
\newcommand{\ZNC}          {\rm{ZNC}\xspace}
\newcommand{\SPD}          {\rm{SPD}\xspace}
\newcommand{\SDD}          {\rm{SDD}\xspace}
\newcommand{\SSD}          {\rm{SSD}\xspace}
\newcommand{\TPC}          {\rm{TPC}\xspace}
\newcommand{\TRD}          {\rm{TRD}\xspace}
\newcommand{\VZERO}        {\rm{V0}\xspace}
\newcommand{\VZEROA}       {\rm{V0A}\xspace}
\newcommand{\VZEROC}       {\rm{V0C}\xspace}
\newcommand{\Vdecay} 	   {\ensuremath{V^{0}}\xspace}

\newcommand{\ee}           {\ensuremath{e^{+}e^{-}}} 
\newcommand{\pip}          {\ensuremath{\pi^{+}}\xspace}
\newcommand{\pim}          {\ensuremath{\pi^{-}}\xspace}
\newcommand{\kap}          {\ensuremath{\rm{K}^{+}}\xspace}
\newcommand{\kam}          {\ensuremath{\rm{K}^{-}}\xspace}
\newcommand{\pbar}         {\ensuremath{\rm\overline{p}}\xspace}
\newcommand{\kzero}        {\ensuremath{{\rm K}^{0}_{\rm{S}}}\xspace}
\newcommand{\lmb}          {\ensuremath{\Lambda}\xspace}
\newcommand{\almb}         {\ensuremath{\overline{\Lambda}}\xspace}
\newcommand{\Om}           {\ensuremath{\Omega^-}\xspace}
\newcommand{\Mo}           {\ensuremath{\overline{\Omega}^+}\xspace}
\newcommand{\X}            {\ensuremath{\Xi^-}\xspace}
\newcommand{\Ix}           {\ensuremath{\overline{\Xi}^+}\xspace}
\newcommand{\Xis}          {\ensuremath{\Xi^{\pm}}\xspace}
\newcommand{\Oms}          {\ensuremath{\Omega^{\pm}}\xspace}
\newcommand{\degree}       {\ensuremath{^{\rm o}}\xspace}
\newcommand{\KK}            {\ensuremath{\rm{K}^{+}\rm{K}^{-}}\xspace}
\newcommand{\pipi}            {\ensuremath{\pi^{+}\pi^{-}}\xspace}

\newcommand{\ie}{\textit{i}.\textit{e}., }
\begin{titlepage}
\PHyear{2023}       
\PHnumber{248}      
\PHdate{03 November}  

\title{Photoproduction of \KK pairs in ultra-peripheral collisions}
\ShortTitle{\KK photoproduction}   

\Collaboration{ALICE Collaboration\thanks{See Appendix~\ref{app:collab} for the list of collaboration members}}
\ShortAuthor{ALICE Collaboration} 

\begin{abstract}

\KK pairs may be produced in photonuclear collisions, either from the decays of photoproduced $\phi (1020)$ mesons, or directly as non-resonant \KK pairs. 
Measurements of \KK photoproduction probe the couplings between the $\phi (1020)$ and charged kaons with photons and nuclear targets. The kaon--proton scattering occurs at energies far above those available elsewhere. 
We present the first measurement of coherent photoproduction of \KK pairs on lead ions in ultra-peripheral collisions using the ALICE detector, including the first investigation of direct \KK production.
There is significant \KK production at low transverse momentum, consistent with coherent photoproduction on lead targets.   In the mass range $1.1 < M_{\rm{KK}} < 1.4$ \GeVmass above the $\phi (1020)$ resonance, for rapidity $|y_{\rm{KK}}|<0.8$ and \ptkk $<$ 0.1 \GeVc, the measured coherent photoproduction cross section is  $\ensuremath{\mathrm{d}\sigma/\mathrm{d}y}$ = 3.37 $\pm\ 0.61$\ (stat.) $\pm\  0.15 $\ (syst.) mb.  The center-of-mass energy per nucleon of the photon--nucleus (Pb) system $W_{\gamma \mathrm{Pb, n}}$ ranges from 33 to 188 GeV, far higher than previous measurements on heavy-nucleus targets. The cross section is larger than expected for $\phi (1020)$ photoproduction alone.   The mass spectrum is fit to a cocktail consisting of $\phi (1020)$ decays, direct \KK photoproduction, and interference between the two.  The confidence regions for the amplitude and relative phase angle for direct \KK photoproduction are presented. 
\end{abstract}
\end{titlepage} 

\setcounter{page}{2} 

{\bf Introduction.} High-energy photoproduction is an important technique for studying hadronic interactions.   Ultra-peripheral collisions (UPCs) of relativistic ions are a tool for studying photonuclear interactions at energies far higher than those available elsewhere~\cite{Bertulani:2005ru,Baltz:2007kq,Contreras:2015dqa,Klein:2020fmr}.  The electromagnetic field of one nucleus forms an intense virtual-photon beam that can interact with nuclei from the opposing beam.  UPC interactions occur when the impact parameter ($b$) between the two nuclei is large enough, e.g. $b$ is greater than the sum of the nuclear radii, so that no obscuring hadron--hadron interactions occur.  

A photon can fluctuate into a quark-antiquark pair (dipole) that scatters elastically from a target nucleus, emerging as a real vector meson~\cite{Klein:2019qfb}. The elastic scattering is mediated by the Pomeron, which is a colorless object and to lowest order, composed of two gluons. The exchange involves the quantum numbers of the vacuum, so following the Vector Meson Dominance (VMD) model, the outgoing meson has the same quantum numbers $J^{PC}=1^{--}$ as the incident photon~\cite{Bauer:1977iq}.  Alternatively, the photon can fluctuate directly into a virtual meson pair, like $\pi^+\pi^-$ or \KK.  One of the mesons can then scatter elastically from the target, making the pair real.  For midrapidity kaons in ALICE, the kaon--proton center of mass energy is 50 GeV, far higher than can be studied elsewhere. Exclusive $K^+K^-$ production can also occur via two-photon~\cite{ALEPH:2003vit} or in double-Pomeron interactions~\cite{STAR:2020dzd}, but this is the first observation in the photon--Pomeron channel. 

Since the channels are indistinguishable, meson pairs from the decay of vector mesons ($\rho^0 \rightarrow \pi^{+}\pi^{-}$ or $\phi (1020) \rightarrow$ \KK) can interfere with the directly produced pairs.  The production amplitude has two terms: the resonance is described using a Breit-Wigner distribution expressed in the Jackson form, with amplitude $A_{\phi}$, and there is, in addition, a continuum component with amplitude $B_{\rm{KK}}$~\cite{Soding:1965nh,Jackson:1964zd,Lebiedowicz:2018eui}, giving
\begin{equation}
\frac{d\sigma}{dM_{\rm{KK}}} = \bigg| A_{\phi}
\frac{\sqrt{M_{\rm{KK}}M_{\phi}\Gamma_\phi}}
{M_{\rm{KK}}^2-M_{\phi}^2+iM_\phi\Gamma_\phi}+B_{\rm{KK}}\bigg|^2,
\label{eq:BW}
\end{equation}
where $M_{\phi}= 1019.416 \pm 0.016$ \MeVmass~\cite{Workman:2022ynf} and $\Gamma_{\phi}$ are the $\phi (1020)$ mass and mass-dependent width, respectively, with
\begin{equation}
    \Gamma_{\phi}= \Gamma_0 \frac{M_{\phi}}{M_{\rm{KK}}} 
    \bigg(\frac{M_{\rm{KK}}^2-4M_{\rm{K}}^2}{M_\phi^2-4M_{\rm{K}}^2}\bigg)^{3/2}.
\end{equation}
Here, $\Gamma_0 = 4.249 \pm 0.013$ \MeVmass is the native \mbox{$\phi$(1020)} width, and $M_{\rm{K}} = 493.677 \pm 0.016$ \MeVmass is the kaon mass~\cite{Workman:2022ynf}.  Both $A_{\phi}$ and $B_{\rm{KK}}$ are complex, but only their relative phase matters. By taking $A_{\phi}$ to be real, the relative phase is encoded in $B_{\rm{KK}}$.  Far above the $\phi (1020)$ resonance (many  $\Gamma_0$), the mass-dependent width rises, and the cross section declines smoothly.  One difference between the \KK and $\pi^+\pi^-$ systems is that the branching ratio $\phi (1020)\rightarrow K^+K^-$ is only 49.1\%~\cite{Workman:2022ynf}, while the $\rho^0$ almost always decays to $\pi^+\pi^-$.  This branching ratio is included in $A_{\phi}$. Ryskin and Shabelski considered the direct dikaon contribution to the total dikaon cross section, and concluded that it should be small and with a relative phase angle near zero~\cite{Ryskin:1997rh}.  A later calculation predicted that the dikaon system should behave similarly to the dipions, with a small correction to the width to account for three-body $\phi (1020)$ decays~\cite{Lebiedowicz:2018eui}. 

The transverse momentum \pt of the meson depends on the production mechanism, so it is important in selecting coherent photoproduction events, where a dipole or virtual meson pair scatters from the target nucleus.   The meson \pt is the vector sum of the photon \pt and the Pomeron \pt, which usually dominates~\cite{Klein:1999gv}.  For coherent production, its scale is controlled by the form factor of the target nucleus.  Destructive interference between the amplitudes for production on the two nuclei can reduce the cross section at low \pt, especially near midrapidity~\cite{Klein:1999gv,STAR:2008llz}.  In incoherent production, when a dipole or virtual meson pair scatters from a single nucleon, the typical \pt is larger, around a few hundred \MeVc.  Although Pomeron--Pomeron interactions can also produce exclusive $K^+K^-$ pairs~\cite{STAR:2020dzd}, because of the short range of the strong force, these reactions cannot be coherent over the entire nuclei, and, so, will not produce a peak at low $p_T$ and do not contribute to the current coherent measurement. 

Previously, $\phi (1020)$ photoproduction has been studied at fixed-target experiments
\cite{Bauer:1977iq,Egloff:1979mg}
and the HERA ep collider~\cite{ZEUS:1996esk,ZEUS:2005bhf,H1:2009cml}.   However, the direct \KK contribution has not yet been observed.   The H1 Collaboration searched for skewing in the $\phi$ Breit-Wigner peak due to direct \KK production (in electroproduction), but found no evidence for it~\cite{H1:2009cml}.  
In contrast, the $\rho^0 + {\rm direct}\ \pi^+\pi^-$ state
has been studied in both UPCs~\cite{STAR:2002caw,STAR:2007elq,STAR:2017enh,ALICE:2020ugp,ALICE:2021jnv,CMS:2019awk} and, at lower energy, at fixed-target experiments~\cite{Bauer:1977iq}.  
The \pipi mass spectra are well fit by the sum of amplitudes for $\rho^0$ and direct $\pi^{+}\pi^{-}$, with a high-statistics fit exhibiting additional interference from $\omega\rightarrow \pi^+\pi^-$~\cite{H1:2020lzc,STAR:2017enh}.  Higher-mass \pipi states have also been seen~\cite{ZEUS:2011tzw,Klein:2016dtn,ALICE:2020ugp}.  

This Letter reports on exclusive photoproduction of the \KK final state, from the decay of the $\phi (1020)$ and direct production. The data cover the mass region $1.1 < M_{\rm{KK}} < 1.4$ \GeVmass.  This is significantly above the $\phi (1020)$ peak, with the lower mass limit at about $M_{\phi}+ 18\Gamma_0$.  

The cross sections are measured at the center-of-mass energy per nucleon of the photon--nucleus (Pb) system $W_{\gamma \mathrm{Pb, n}}$, where $W_{\gamma \mathrm{Pb, n}}$ varies from 33 to 188 \GeV, depending on $M_{\rm{KK}}$ and rapidity.  This range of energies is more than an order of magnitude higher than previous studies using heavy-nucleus targets~\cite{Bauer:1977iq,Egloff:1979mg}.

{\bf Detector and data.}  The results presented in this Letter are based on the data collected in 2015 by the ALICE experiment~\cite{ALICE:2008aecl,ALICE:2014sbx}, with Pb--Pb collisions at a center-of-mass energy per nucleon pair \fivenn. 
A dedicated trigger was used to select candidate UPC events~\cite{ALICE:2020ugp}, rejecting any activity within the time windows for
nominal beam--beam interactions, using the scintillator detectors \rm{V0}\xspace~\cite{ALICE:2013axi} and \rm{AD}\xspace~\cite{LHCForwardPhysicsWorkingGroup:2016ote,Broz:2020ejr} located at large positive and negative pseudorapidity. 
In addition, the trigger required that the Silicon Pixel Detector (SPD), the two innermost layers of the inner tracking system (ITS)~\cite{ALICE:2010tia}, measured at least two short track segments  with a large opening angle in azimuth.

The time projection chamber (\TPC) covering the pseudorapidity acceptance of \etarange 0.9 
was used for charged particle tracking and vertexing together with the \ITS  as well as for particle identification based on the specific ionization energy loss, \dEdx ~\cite{Alme:2010ke}.

{\bf Analysis Procedure.} The analysis selected events with exactly two good tracks. The tracks were required to have at least 50 hits (clusters) in the TPC, at least one hit in each of the two layers of the SPD, and to have the distance-of-closest approach to the event vertex of less than $0.0182 + 0.035/\pt^{1.01}$ cm in the transverse plane and less than 2 cm along the beam direction. The two selected tracks having opposite charge are reconstructed as \KK pair candidates under the kaon mass hypothesis. As there are no same-charge pairs passing the particle identification criteria, the contribution of uncorrelated background could be ignored in this analysis.

Kaons are identified based on the \dEdx measurement in the \TPC. The selection criteria are applied to the variable $n_{\sigma_{i}}$, the deviation of the measured signal from the expected signal in units of the \dEdx measurement uncertainty for each particle hypothesis $i$. Since the ratio of signal \KK pairs to background \pipi pairs is less than 0.1\%, stringent particle identification criteria are introduced. First, the tracks in each pair are required to satisfy $|n_{\sigma_{\rm K}}| < 3$. In addition, the tracks which are compatible within $2 n_{\sigma_{\pi, \mu, e}}$ are excluded to reject \pipi pairs as well as dilepton pairs from the $\gamma\gamma\rightarrow l^{+}l^{-}$ process. 

The contamination of the \KK pair candidates by the misidentified particles is estimated from the two-dimensional $n_{\sigma_K}$ distribution of the two tracks in each pair. While the signal \KK pairs have a two-dimensional Gaussian-like distribution centered at (0,0), background pairs are clustered at non-zero values. For 1.1 $< M_{\rm{KK}} <$ 1.4 \GeVmass, the contamination is negligible as the signal and background distributions are well separated from each other. At higher masses, the expected difference in \dEdx between kaons and lighter particles decreases, so they become indistinguishable in some $n_{\sigma_K}$ regions. Therefore, the invariant mass range above 1.4 \GeVmass is not used in the analysis.  For pairs with $M_{\rm{KK}} <1.1$ \GeVmass, the kaons lose energy rapidly and do not reach the sensitive region of the ALICE detector.

{\bf Measurement of the cross section.} The invariant mass differential cross section of exclusive \KK photoproduction is obtained by correcting the number of \KK candidates ($N_{\rm{KK}}$) found in  the rapidity interval of $|y_{\rm{KK}}|<0.8$ and in the \pt interval of $\ptkk < 0.1 $\ \GeVc by acceptance and efficiency ($\mathcal{A} \times \epsilon$),
\begin{equation}
    \frac{\mathrm{d}^{2}\it{\sigma}}{\mathrm{d}\it{M}_{\rm{KK}}\mathrm{d}y_{\rm{KK}}} = \frac{\it{N}_{\rm{KK}}\times f_{\mathrm{pileup}}}{(\mathcal{A} \times \epsilon)\times \mathcal{L}\times \Delta M_{\rm{KK}}\times \Delta y_{\rm{KK}}}.
\end{equation}
The $\mathcal{A} \times \epsilon$ is computed using a dedicated Monte Carlo simulation with STARlight~\cite{Klein:2016yzr} for the \KK pairs from direct production and $\phi (1020)$ decays. 

The generated \KK pairs are transported through the detector setup using a GEANT\ 3 model~\cite{Brun:1082634} to simulate a realistic detector response. The corresponding integrated luminosity ($\mathcal{L}$) for the data sample is 0.406 $\mu \rm{b}^{-1}$ with a relative systematic uncertainty of 2.6\%~\cite{ALICE:2022xir}. 
Some events are lost due to pileup, when another interaction creates a signal in one of the veto detectors.
The pileup events mainly come from two-photon production of $\rm{e}^+\rm{e}^-$ pairs, and their effect is taken into account with an additional correction factor, \mbox{$f_{\mathrm{pileup}}$ = 11.1 $\pm$ 3.8\%~\cite{ALICE:2020ugp}}. Similarly, the \ptsquared-differential cross section of exclusive \KK photoproduction in 1.1 $< M_{\rm{KK}} <$ 1.4 \GeVmass is measured,
\begin{equation}
    \frac{\mathrm{d}^{2}\it{\sigma}}{\mathrm{d}\ptsquaredkk\mathrm{d}\it{y}_{\rm{KK}}} = \frac{\it{N}_{\rm{KK}}\times f_{\mathrm{pileup}}}{(\mathcal{A} \times \epsilon)\times \mathcal{L}\times \Delta \ptsquaredkk \times \Delta y_{\rm{KK}}}.
\end{equation}

{\bf Systematic uncertainties.} The systematic uncertainties of the measured cross section are estimated for the track selection criteria (1.5\%) and the track matching between \ITS and \TPC (4\%) as well as for the acceptance and efficiency (1\%), without dependence on \ptkk, $y_{\rm{KK}}$, and $M_{\rm{KK}}$~\cite{ALICE:2020ugp}. Uncertainties of 1\% and 3.8\% are included for the trigger efficiency and for the pileup correction, respectively~\cite{ALICE:2020ugp}. The uncertainty of the luminosity (2.6\%) results from the uncertainty of the reference luminosity in the cross section of 2.5\%~\cite{ALICE:2022xir} and an additional 0.4\% uncertainty on the live-time of readout detectors used for the trigger.

The systematic uncertainty of the kaon identification is estimated as a function of \ptkk and $M_{\rm{KK}}$ to account for the track momentum dependence in the kaon identification performance. The expected signal of \TPC \dEdx for each particle hypothesis is varied in the MC simulations by the maximum difference of the signal in data and MC simulations. Then, $n_{\sigma_{i}}$ and the corresponding kaon identification efficiency are recalculated. The resulting uncertainty is negligible for 1.1 $< M_{\rm{KK}} <$ 1.2 \GeVmass and amounts to 3.9\% and 6.5\% for 1.2 $< M_{\rm{KK}} <$ 1.3 \GeVmass and 1.3 $< M_{\rm{KK}} <$ 1.4 \GeVmass, respectively. The systematic uncertainty increases slowly as a function of \ptkk, from 3.4\% for \ptkk $<$ 0.025 \GeVc to 4.9\% for \mbox{0.1 $< \ptkk <$ 0.2 \GeVc.}  Over most of the kinematic range, this is the largest single systematic uncertainty.

\begin{figure}[!htbp]
    \centering
    \includegraphics[width=0.65\textwidth]{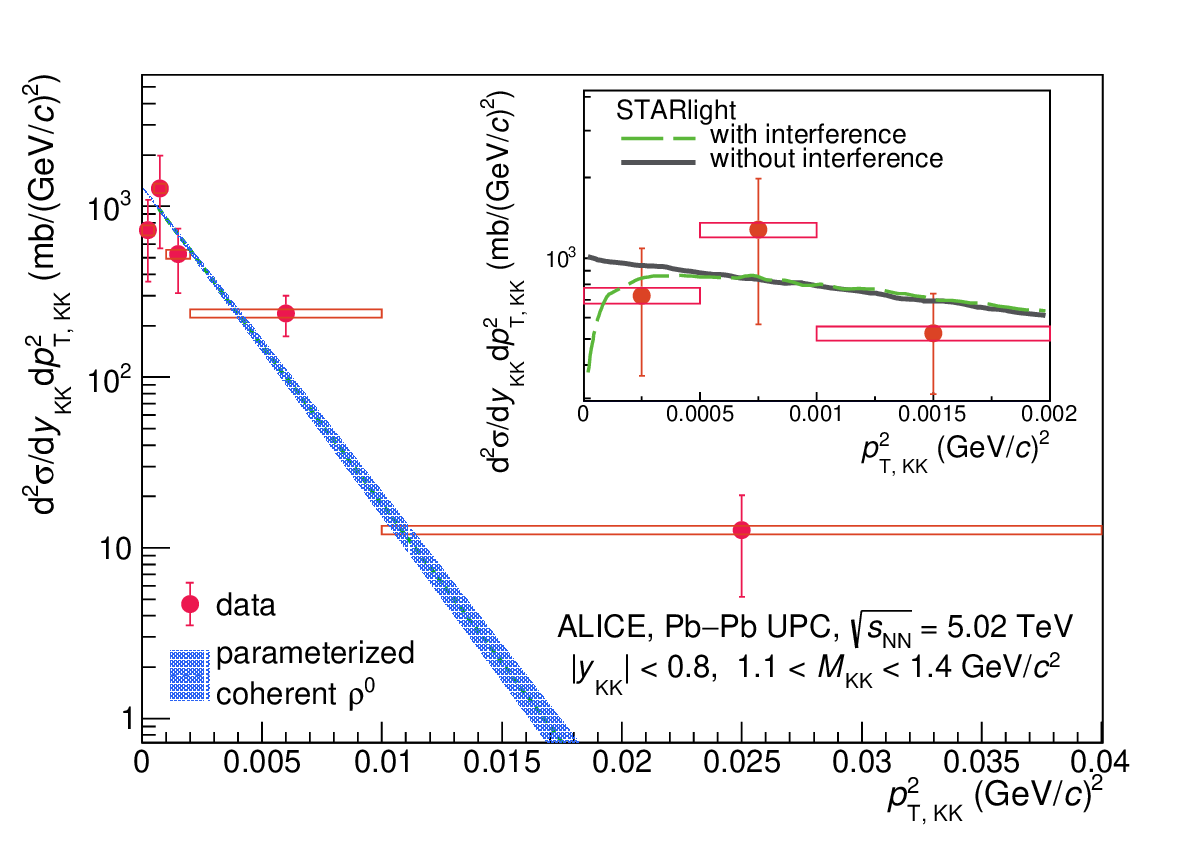} 
    \caption{Differential cross section as a function of $\it{p}_{\rm{T, KK}}^{\rm{2}}$ for exclusive \KK photoproduction in Pb--Pb UPCs at \mbox{$\sqrt{s_{\rm{NN}}}$= 5.02 \TeV} and $|y_{\rm{KK}}|<0.8$. The vertical lines and boxes across the data points represent statistical and systematic uncertainties, respectively. The dashed blue line and band are the result of a fit to an exponential with the fixed slope parameter \mbox{$b=428\ \pm\ 6\ ({\rm stat.})\ \pm\ 15\ ({\rm syst.})$ \inverseGeVcSquared,} from a previous result on $\rho^{0}$ production~\cite{ALICE:2015nbw} (see the text for details). The inset shows two curves from STARlight with and without interference between the two photon directions~\cite{Klein:1999gv,Klein:2016yzr}.}
\label{fig:pt2}
\end{figure}

{\bf Results.} 
Figure~\ref{fig:pt2} shows the \ptsquared spectrum of the selected \KK events. Most of the cross section is concentrated below \ptsquaredkk $<$ 0.01 \GeVcSquared, consistent with coherent photoproduction.  Some events are seen at higher values of \ptsquaredkk; these may be from incoherent production.  The coherent data are well described with an 
exponential shape $\rm{d}^2\it{\sigma}/\rm{d}\it{y}\rm{d}\it{p}_{\rm{T}}^{\rm{2}} = a \exp(-b\pt^{\rm{2}})$,
where the slope parameter $b$ is fixed to that measured for coherent $\rho^0$ photoproduction on lead, \mbox{$~{b = 428\ \pm\ 6\ ({\rm stat.})\ \pm\ 15\ ({\rm syst.})}$ \inverseGeVcSquared}~\cite{ALICE:2015nbw}.  The figure inset shows an expanded view of the low \ptsquared region, and compares the data with two STARlight calculations, with and without an interference between photon emission from the two nuclei~\cite{Klein:1999gv,Klein:2016yzr}, which exhibit different trends only at very low \pt. The curve with interference is a slightly better match to the data.  

\begin{figure}[!hbtp]
    \centering
    \includegraphics[width=0.65\textwidth]{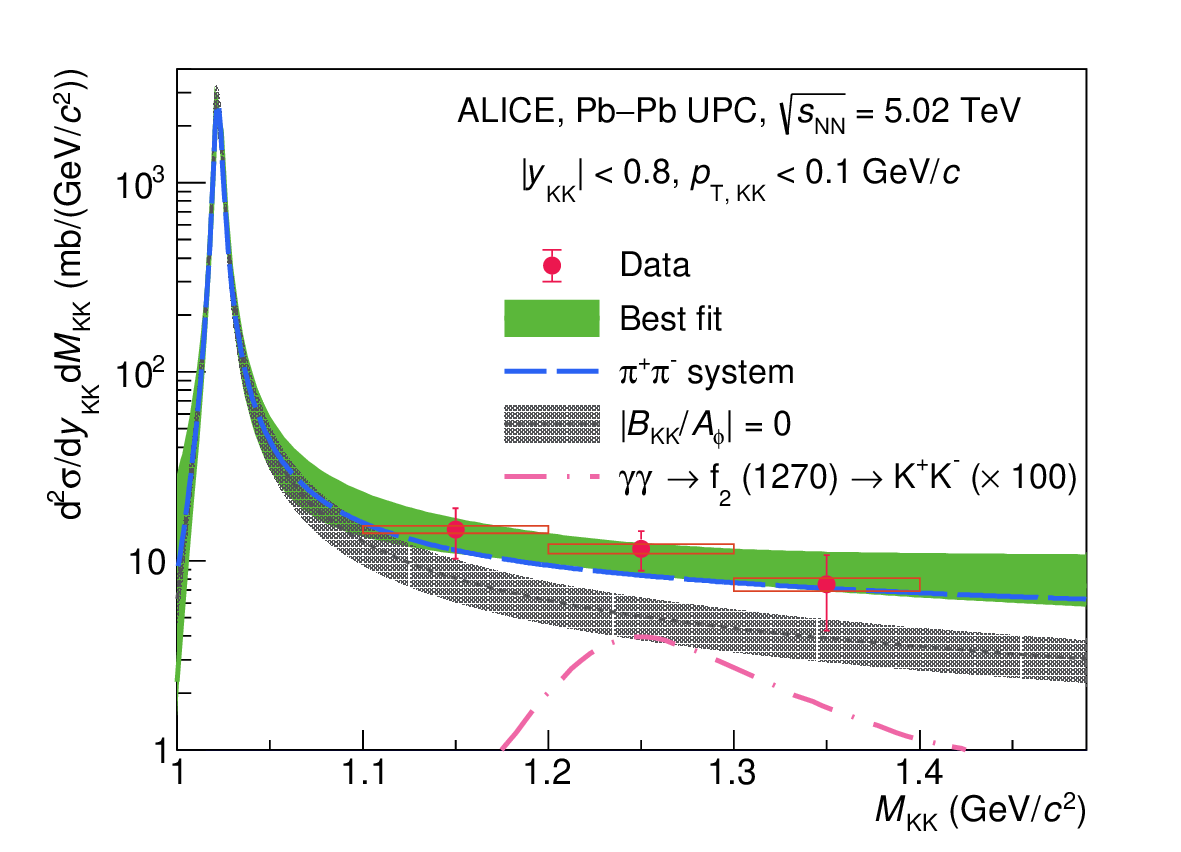} 
    \caption{Differential cross section of coherent \KK photoproduction as a function $\it{M_\mathrm{KK}}$ in Pb--Pb UPCs at \mbox{\snn = 5.02 \TeV} in $|y_\mathrm{KK}|\ <\ 0.8$. The vertical lines and boxes along the data points represent statistical and systematic uncertainties, respectively. The green solid line presents the best fit result of  \mbox{$|B_{\rm{KK}}/A_{\phi}|=0.28$ \GeVmassSR} and $\varphi$ = 0.06 rad together with the 1$\sigma$ bounds of the fit in a green band. The blue dashed curve shows the best fit with $|B_{\pi\pi}/A_{\rho}|=0.54$ \GeVmassSR~\cite{ALICE:2020ugp}  and $\varphi$ = 1.46 rad (the best-fit values for $\rho$ plus direct $\pi^+\pi^-$)~\cite{STAR:2017enh}. The black dotted line represents the curves under the hypothesis of $|B_{\rm{KK}}/A_{\phi}|=0$, showing only the $\phi  (1020)\rightarrow$ \KK contribution. The gray band indicates the impact of the systematic uncertainty from the $\phi (1020)$ meson cross section, showing a 25$\%$ variation.}
\label{fig:mKK}
\end{figure}

The invariant mass dependent cross section for coherent \KK photoproduction is shown in Fig.~\ref{fig:mKK}. The data with \ptsquaredkk $>$ 0.01 \GeVcSquared are mostly from incoherent photoproduction, and so are not included in the cross sections. The integrated cross section $\ensuremath{\mathrm{d}\sigma/\mathrm{d}y_{\rm{KK}}}$ = 3.37 $\pm\ 0.61$\ (stat.)\ $\pm\ 0.15 $\ (syst.) mb is measured in the mass range $1.1 < M_{\rm{KK}} < 1.4$ \GeVmass for rapidity $|y_{\rm{KK}}|<0.8$ and \mbox{\ptsquaredkk $<$ 0.01 \GeVcSquared}.   \KK pairs could be produced by other reactions, such as $\gamma\gamma\rightarrow f_2 (1270)\rightarrow$ \KK, but calculations indicate that the expected cross section for this process~\cite{Baltz:2009jk} estimated using STARlight~\cite{Klein:1999qj} is a negligible fraction of the \KK photoproduction cross section as illustrated in Fig.~\ref{fig:mKK}.

The measured cross section is fitted to a combination of $\phi  (1020)\rightarrow$ \KK and direct \KK production according to Eq.~(\ref{eq:BW}).  The amplitude of $\phi (1020)\rightarrow$ \KK ($A_{\phi}$) was calculated from previous photoproduction measurements on protons~\cite{Egloff:1979mg,ZEUS:1996esk,Crittenden:1997yz} and a Glauber calculation~\cite{Klein:1999qj}, with a branching ratio of $\phi (1020)\rightarrow \KK ( 49.2\ \pm\ 0.5 \%$)~\cite{Workman:2022ynf}.  The reference $\phi (1020)$ cross sections did not include a direct \KK contribution.  This could have had a small effect on the measured $A_{\phi}$.

A 25\% uncertainty in $A_{\phi}$ is estimated to account for the uncertainty in the previous measurements~\cite{Egloff:1979mg,ZEUS:1996esk} and the uncertainty in the Glauber approach~\cite{Frankfurt:2015cwa}. STARlight predictions for $\rho^0$ production under similar circumstances were 15\%--20\% below the data~\cite{ALICE:2020ugp}, while predictions for $J/\psi$ production were about 15\%--50\% above the data, depending on rapidity~\cite{ALICE:2021gpt}.  The latter is not surprising, since STARlight does not include gluon shadowing.  The $\phi (1020)$ is intermediate in mass, but closer to the $\rho^0$, so a $\pm 25\%$ uncertainty in the cross section seems conservative for the $\phi (1020)$. 

The black dotted line and surrounding shaded region in Fig.~\ref{fig:mKK} show the $\phi$-only prediction, with the 25\% uncertainty. The measured cross section is about 2.1 $\sigma$ above the expected $\phi (1020)\rightarrow$ \KK cross section in the range  $1.1 < M_{\rm{KK}} < 1.4$ \GeVmass.  Also shown, with a blue dashed line, is the prediction using the values 
$|B_{\pi\pi}/A_{\rho}|=0.54\ \pm\ 0.01$\ (stat.)\ $\pm\ 0.02 $\ (syst.) \GeVmassSR~\cite{ALICE:2020ugp}  and 
relative phase angle $\varphi = 1.46\ \pm\ 0.11$\ (stat.)\ $\pm\ 0.07 $\ (syst.) rad~\cite{STAR:2017enh} found for $\rho^0$ plus the direct $\pi^+\pi^-$ system. The resulting $\rm{d}^2\sigma/\rm{d}\it{M_{\rm{KK}}}\rm{d}\it{y_{\rm{KK}}}$ is slightly below, but consistent with, the data points.  The best fit of Eq.~(\ref{eq:BW}) found the relative fraction of direct \KK contribution with respect to $A_\phi$ to be $|B_{\rm{KK}}/A_\phi|= 0.28$ \GeVmassSR, while the relative phase angle between $\phi (1020) \rightarrow$ \KK and direct \KK  is 0.06 rad. 

\begin{figure}[!htbp]
    \centering
    \includegraphics[width=0.65\textwidth]{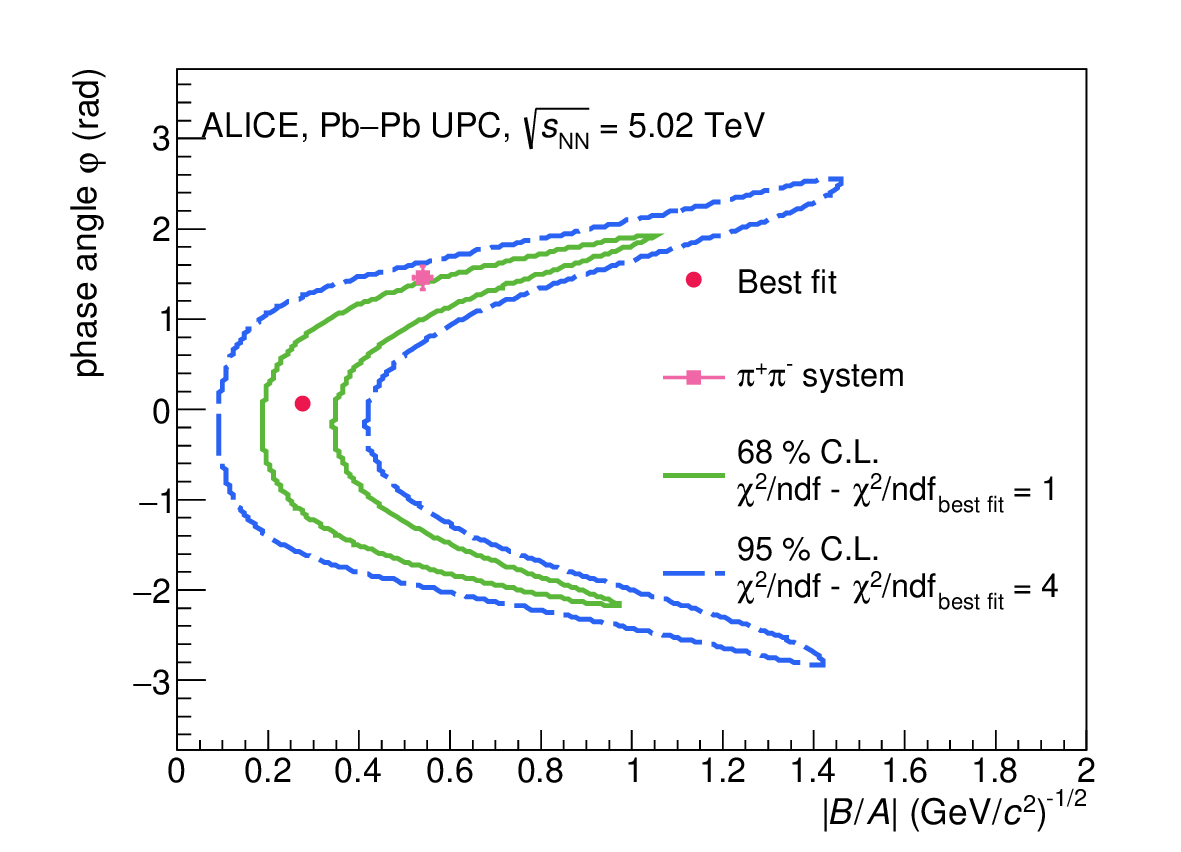} 
    \caption{Confidence regions for the relative fraction of direct \KK contribution with respect to the amplitude of $\phi (1020) \rightarrow$ \KK ($|B_{\rm{KK}}/A_{\phi}|$) and the relative phase angle between $\phi(1020)\rightarrow$\KK and direct \KK ($\varphi$). The best fit is shown as a red dot at $|B_{\rm{KK}}/A_{\phi}|=0.28$ \GeVmassSR and $\varphi = $ 0.06 rad found with $\chi^{2}/\rm{ndf_{best\ fit}} = 0.3$, while a pink square at the $|B_{\pi\pi}/A_{\rho}|=0.54\ \pm\ 0.01$\ (stat.)\ $\pm\ 0.02 $\ (syst.) \GeVmassSR~\cite{ALICE:2020ugp}  and 
relative phase angle $\varphi = 1.46\ \pm\ 0.11$\ (stat.)\ $\pm\ 0.07 $\ (syst.) rad~\cite{STAR:2017enh} indicates the best-fit values for $\rho$ plus direct $\pi^+\pi^-$. The green solid line and blue dashed line represent the boundary of 68\% and 95\% confidence regions, respectively.  }
\label{fig:mKK_par}
\end{figure}

Figure~\ref{fig:mKK_par} shows the 
confidence regions for $|B_{\rm{KK}}/A_{\phi}|$ and $\varphi$. 
The horseshoe shape of the curves including the first investigation of direct $K^+K^-$ production is because of the large correlations between the two parameters.  If the interference is constructive, a small direct \KK component is preferred, while a large \KK component is better fit with destructive interference.   The invariant mass-dependent cross section curves corresponding to the 68$\%$ confidence region are shown as a $1\sigma$ green band in Fig.~\ref{fig:mKK_par}, while the dashed blue band shows the 95\% confidence level.  We do not include the uncertainty on the cross section $\sigma(\gamma A\rightarrow\phi (1020) A)$ within the figure.  As $\sigma(\gamma A\ \rightarrow\ \phi (1020) A)$ is reduced or increased, it moves the confidence region left or right, with relatively small changes to the shape.  One standard deviation ($\pm\ 1\sigma$) changes in $\sigma(\gamma A\rightarrow\phi (1020) A)$ move $|B_{\rm{KK}}/A_{\phi}|$ by about $\pm\ 0.15$
\GeVmassSR respectively.   The best-fit point for the $\pi\pi$ system is fully compatible with the current \KK measurement.

Looking ahead, during LHC Run 3, ALICE will collect a far larger data sample, due to the increased luminosity~\cite{Citron:2018lsq} and continuous-readout data acquisition system~\cite{Buncic:2011297,CERN-LHCC-2013-020}, which eliminates the need for a restrictive, prescaled trigger. Improved precision measurement of coherent $\mathrm{K}^+\mathrm{K}^-$ photoproduction by the reduction of statistical uncertainty and an improved tracking 
will make it possible to further disentangle the resonance and non-resonance contributions with their relative phase angle.

{\bf Conclusions.}  We report the first study of coherent \KK photoproduction in ultra-peripheral collisions at $W_{\gamma \mathrm{Pb, n}}$ from
33 to 188 GeV, in the range $1.1 < M_{\rm{KK}} < 1.4$ \GeVmass and $|y_{\mathrm{KK}}|<0.8$. The $\rm{d}^2\sigma/\rm{d}\it{p_{\rm T, KK}^{2}}\rm{d}\it{y_{\rm{KK}}}$ is concentrated below \ptsquaredkk $ < 0.01 \ $\GeVcSquared, consistent with coherent photoproduction.
The measured $\rm{d}^2\sigma/\rm{d}\it{ M_{\rm{KK}}}\rm{d}\it{y_{\rm{KK}}}$ below \ptsquaredkk $<$ 0.01 \GeVcSquared is about 2.1 $\sigma$ larger than what is expected from $\phi (1020)$ production alone estimated based on HERA data with a Glauber model calculation, but is consistent with a mixture of $\phi (1020)$ and direct \KK production.  The fitted ratio of $\phi (1020)$ production to $\KK$ production is consistent with that seen for the $\rho^0$ and direct $\pi^{+}\pi^{-}$.

\newenvironment{acknowledgement}{\relax}{\relax}
\begin{acknowledgement}
\section*{Acknowledgements}

The ALICE Collaboration would like to thank all its engineers and technicians for their invaluable contributions to the construction of the experiment and the CERN accelerator teams for the outstanding performance of the LHC complex.
The ALICE Collaboration gratefully acknowledges the resources and support provided by all Grid centres and the Worldwide LHC Computing Grid (WLCG) collaboration.
The ALICE Collaboration acknowledges the following funding agencies for their support in building and running the ALICE detector:
A. I. Alikhanyan National Science Laboratory (Yerevan Physics Institute) Foundation (ANSL), State Committee of Science and World Federation of Scientists (WFS), Armenia;
Austrian Academy of Sciences, Austrian Science Fund (FWF): [M 2467-N36] and Nationalstiftung f\"{u}r Forschung, Technologie und Entwicklung, Austria;
Ministry of Communications and High Technologies, National Nuclear Research Center, Azerbaijan;
Conselho Nacional de Desenvolvimento Cient\'{\i}fico e Tecnol\'{o}gico (CNPq), Financiadora de Estudos e Projetos (Finep), Funda\c{c}\~{a}o de Amparo \`{a} Pesquisa do Estado de S\~{a}o Paulo (FAPESP) and Universidade Federal do Rio Grande do Sul (UFRGS), Brazil;
Bulgarian Ministry of Education and Science, within the National Roadmap for Research Infrastructures 2020-2027 (object CERN), Bulgaria;
Ministry of Education of China (MOEC) , Ministry of Science \& Technology of China (MSTC) and National Natural Science Foundation of China (NSFC), China;
Ministry of Science and Education and Croatian Science Foundation, Croatia;
Centro de Aplicaciones Tecnol\'{o}gicas y Desarrollo Nuclear (CEADEN), Cubaenerg\'{\i}a, Cuba;
Ministry of Education, Youth and Sports of the Czech Republic, Czech Republic;
The Danish Council for Independent Research | Natural Sciences, the VILLUM FONDEN and Danish National Research Foundation (DNRF), Denmark;
Helsinki Institute of Physics (HIP), Finland;
Commissariat \`{a} l'Energie Atomique (CEA) and Institut National de Physique Nucl\'{e}aire et de Physique des Particules (IN2P3) and Centre National de la Recherche Scientifique (CNRS), France;
Bundesministerium f\"{u}r Bildung und Forschung (BMBF) and GSI Helmholtzzentrum f\"{u}r Schwerionenforschung GmbH, Germany;
General Secretariat for Research and Technology, Ministry of Education, Research and Religions, Greece;
National Research, Development and Innovation Office, Hungary;
Department of Atomic Energy Government of India (DAE), Department of Science and Technology, Government of India (DST), University Grants Commission, Government of India (UGC) and Council of Scientific and Industrial Research (CSIR), India;
National Research and Innovation Agency - BRIN, Indonesia;
Istituto Nazionale di Fisica Nucleare (INFN), Italy;
Japanese Ministry of Education, Culture, Sports, Science and Technology (MEXT) and Japan Society for the Promotion of Science (JSPS) KAKENHI, Japan;
Consejo Nacional de Ciencia (CONACYT) y Tecnolog\'{i}a, through Fondo de Cooperaci\'{o}n Internacional en Ciencia y Tecnolog\'{i}a (FONCICYT) and Direcci\'{o}n General de Asuntos del Personal Academico (DGAPA), Mexico;
Nederlandse Organisatie voor Wetenschappelijk Onderzoek (NWO), Netherlands;
The Research Council of Norway, Norway;
Commission on Science and Technology for Sustainable Development in the South (COMSATS), Pakistan;
Pontificia Universidad Cat\'{o}lica del Per\'{u}, Peru;
Ministry of Education and Science, National Science Centre and WUT ID-UB, Poland;
Korea Institute of Science and Technology Information and National Research Foundation of Korea (NRF), Republic of Korea;
Ministry of Education and Scientific Research, Institute of Atomic Physics, Ministry of Research and Innovation and Institute of Atomic Physics and Universitatea Nationala de Stiinta si Tehnologie Politehnica Bucuresti, Romania;
Ministry of Education, Science, Research and Sport of the Slovak Republic, Slovakia;
National Research Foundation of South Africa, South Africa;
Swedish Research Council (VR) and Knut \& Alice Wallenberg Foundation (KAW), Sweden;
European Organization for Nuclear Research, Switzerland;
Suranaree University of Technology (SUT), National Science and Technology Development Agency (NSTDA) and National Science, Research and Innovation Fund (NSRF via PMU-B B05F650021), Thailand;
Turkish Energy, Nuclear and Mineral Research Agency (TENMAK), Turkey;
National Academy of  Sciences of Ukraine, Ukraine;
Science and Technology Facilities Council (STFC), United Kingdom;
National Science Foundation of the United States of America (NSF) and United States Department of Energy, Office of Nuclear Physics (DOE NP), United States of America.
In addition, individual groups or members have received support from:
Czech Science Foundation (grant no. 23-07499S), Czech Republic;
European Research Council, Strong 2020 - Horizon 2020 (grant nos. 950692, 824093), European Union;
ICSC - Centro Nazionale di Ricerca in High Performance Computing, Big Data and Quantum Computing, European Union - NextGenerationEU;
Academy of Finland (Center of Excellence in Quark Matter) (grant nos. 346327, 346328), Finland.

\end{acknowledgement}

\bibliographystyle{utphys} 
\bibliography{bibliography}

\newpage
\appendix

%
%

\section{The ALICE Collaboration}
\label{app:collab}
\begin{flushleft} 
\small

S.~Acharya\,\orcidlink{0000-0002-9213-5329}\,$^{\rm 128}$, 
D.~Adamov\'{a}\,\orcidlink{0000-0002-0504-7428}\,$^{\rm 87}$, 
G.~Aglieri Rinella\,\orcidlink{0000-0002-9611-3696}\,$^{\rm 33}$, 
M.~Agnello\,\orcidlink{0000-0002-0760-5075}\,$^{\rm 30}$, 
N.~Agrawal\,\orcidlink{0000-0003-0348-9836}\,$^{\rm 52}$, 
Z.~Ahammed\,\orcidlink{0000-0001-5241-7412}\,$^{\rm 136}$, 
S.~Ahmad\,\orcidlink{0000-0003-0497-5705}\,$^{\rm 16}$, 
S.U.~Ahn\,\orcidlink{0000-0001-8847-489X}\,$^{\rm 72}$, 
I.~Ahuja\,\orcidlink{0000-0002-4417-1392}\,$^{\rm 38}$, 
A.~Akindinov\,\orcidlink{0000-0002-7388-3022}\,$^{\rm 142}$
% ...

\section*{Affiliation Notes}

$^{\rm I}$ Deceased\\
$^{\rm II}$ Also at: Max-Planck-Institut fur Physik, Munich, Germany\\
$^{\rm III}$ Also at: Italian National Agency for New Technologies, Energy and Sustainable Economic Development (ENEA), Bologna, Italy\\
$^{\rm IV}$ Also at: Dipartimento DET del Politecnico di Torino, Turin, Italy\\
$^{\rm V}$ Also at: Yildiz Technical University, Istanbul, T\"{u}rkiye\\
$^{\rm VI}$ Also at: Department of Applied Physics, Aligarh Muslim University, Aligarh, India\\
$^{\rm VII}$ Also at: Institute of Theoretical Physics, University of Wroclaw, Poland\\
$^{\rm VIII}$ Also at: An institution covered by a cooperation agreement with CERN\\

\section*{Collaboration Institutes}

$^{1}$ A.I. Alikhanyan National Science Laboratory (Yerevan Physics Institute) Foundation, Yerevan, Armenia\\
$^{2}$ AGH University of Krakow, Cracow, Poland\\
$^{3}$ Bogolyubov Institute for Theoretical Physics, National Academy of Sciences of Ukraine, Kiev, Ukraine\\
$^{4}$ Bose Institute, Department of Physics  and Centre for Astroparticle Physics and Space Science (CAPSS), Kolkata, India\\
$^{5}$ California Polytechnic State University, San Luis Obispo, California, United States\\
$^{6}$ Central China Normal University, Wuhan, China\\
$^{7}$ Centro de Aplicaciones Tecnol\'{o}gicas y Desarrollo Nuclear (CEADEN), Havana, Cuba\\
$^{8}$ Centro de Investigaci\'{o}n y de Estudios Avanzados (CINVESTAV), Mexico City and M\'{e}rida, Mexico\\
$^{9}$ Chicago State University, Chicago, Illinois, United States\\
$^{10}$ China Institute of Atomic Energy, Beijing, China\\
$^{11}$ China University of Geosciences, Wuhan, China\\
$^{12}$ Chungbuk National University, Cheongju, Republic of Korea\\
$^{13}$ Comenius University Bratislava, Faculty of Mathematics, Physics and Informatics, Bratislava, Slovak Republic\\
$^{14}$ COMSATS University Islamabad, Islamabad, Pakistan\\
$^{15}$ Creighton University, Omaha, Nebraska, United States\\
$^{16}$ Department of Physics, Aligarh Muslim University, Aligarh, India\\
$^{17}$ Department of Physics, Pusan National University, Pusan, Republic of Korea\\
$^{18}$ Department of Physics, Sejong University, Seoul, Republic of Korea\\
$^{19}$ Department of Physics, University of California, Berkeley, California, United States\\
$^{20}$ Department of Physics, University of Oslo, Oslo, Norway\\
$^{21}$ Department of Physics and Technology, University of Bergen, Bergen, Norway\\
$^{22}$ Dipartimento di Fisica, Universit\`{a} di Pavia, Pavia, Italy\\
$^{23}$ Dipartimento di Fisica dell'Universit\`{a} and Sezione INFN, Cagliari, Italy\\
$^{24}$ Dipartimento di Fisica dell'Universit\`{a} and Sezione INFN, Trieste, Italy\\
$^{25}$ Dipartimento di Fisica dell'Universit\`{a} and Sezione INFN, Turin, Italy\\
$^{26}$ Dipartimento di Fisica e Astronomia dell'Universit\`{a} and Sezione INFN, Bologna, Italy\\
$^{27}$ Dipartimento di Fisica e Astronomia dell'Universit\`{a} and Sezione INFN, Catania, Italy\\
$^{28}$ Dipartimento di Fisica e Astronomia dell'Universit\`{a} and Sezione INFN, Padova, Italy\\
$^{29}$ Dipartimento di Fisica `E.R.~Caianiello' dell'Universit\`{a} and Gruppo Collegato INFN, Salerno, Italy\\
$^{30}$ Dipartimento DISAT del Politecnico and Sezione INFN, Turin, Italy\\
$^{31}$ Dipartimento di Scienze MIFT, Universit\`{a} di Messina, Messina, Italy\\
$^{32}$ Dipartimento Interateneo di Fisica `M.~Merlin' and Sezione INFN, Bari, Italy\\
$^{33}$ European Organization for Nuclear Research (CERN), Geneva, Switzerland\\
$^{34}$ Faculty of Electrical Engineering, Mechanical Engineering and Naval Architecture, University of Split, Split, Croatia\\
$^{35}$ Faculty of Engineering and Science, Western Norway University of Applied Sciences, Bergen, Norway\\
$^{36}$ Faculty of Nuclear Sciences and Physical Engineering, Czech Technical University in Prague, Prague, Czech Republic\\
$^{37}$ Faculty of Physics, Sofia University, Sofia, Bulgaria\\
$^{38}$ Faculty of Science, P.J.~\v{S}af\'{a}rik University, Ko\v{s}ice, Slovak Republic\\
$^{39}$ Frankfurt Institute for Advanced Studies, Johann Wolfgang Goethe-Universit\"{a}t Frankfurt, Frankfurt, Germany\\
$^{40}$ Fudan University, Shanghai, China\\
$^{41}$ Gangneung-Wonju National University, Gangneung, Republic of Korea\\
$^{42}$ Gauhati University, Department of Physics, Guwahati, India\\
$^{43}$ Helmholtz-Institut f\"{u}r Strahlen- und Kernphysik, Rheinische Friedrich-Wilhelms-Universit\"{a}t Bonn, Bonn, Germany\\
$^{44}$ Helsinki Institute of Physics (HIP), Helsinki, Finland\\
$^{45}$ High Energy Physics Group,  Universidad Aut\'{o}noma de Puebla, Puebla, Mexico\\
$^{46}$ Horia Hulubei National Institute of Physics and Nuclear Engineering, Bucharest, Romania\\
$^{47}$ HUN-REN Wigner Research Centre for Physics, Budapest, Hungary\\
$^{48}$ Indian Institute of Technology Bombay (IIT), Mumbai, India\\
$^{49}$ Indian Institute of Technology Indore, Indore, India\\
$^{50}$ INFN, Laboratori Nazionali di Frascati, Frascati, Italy\\
$^{51}$ INFN, Sezione di Bari, Bari, Italy\\
$^{52}$ INFN, Sezione di Bologna, Bologna, Italy\\
$^{53}$ INFN, Sezione di Cagliari, Cagliari, Italy\\
$^{54}$ INFN, Sezione di Catania, Catania, Italy\\
$^{55}$ INFN, Sezione di Padova, Padova, Italy\\
$^{56}$ INFN, Sezione di Pavia, Pavia, Italy\\
$^{57}$ INFN, Sezione di Torino, Turin, Italy\\
$^{58}$ INFN, Sezione di Trieste, Trieste, Italy\\
$^{59}$ Inha University, Incheon, Republic of Korea\\
$^{60}$ Institute for Gravitational and Subatomic Physics (GRASP), Utrecht University/Nikhef, Utrecht, Netherlands\\
$^{61}$ Institute of Experimental Physics, Slovak Academy of Sciences, Ko\v{s}ice, Slovak Republic\\
$^{62}$ Institute of Physics, Homi Bhabha National Institute, Bhubaneswar, India\\
$^{63}$ Institute of Physics of the Czech Academy of Sciences, Prague, Czech Republic\\
$^{64}$ Institute of Space Science (ISS), Bucharest, Romania\\
$^{65}$ Institut f\"{u}r Kernphysik, Johann Wolfgang Goethe-Universit\"{a}t Frankfurt, Frankfurt, Germany\\
$^{66}$ Instituto de Ciencias Nucleares, Universidad Nacional Aut\'{o}noma de M\'{e}xico, Mexico City, Mexico\\
$^{67}$ Instituto de F\'{i}sica, Universidade Federal do Rio Grande do Sul (UFRGS), Porto Alegre, Brazil\\
$^{68}$ Instituto de F\'{\i}sica, Universidad Nacional Aut\'{o}noma de M\'{e}xico, Mexico City, Mexico\\
$^{69}$ iThemba LABS, National Research Foundation, Somerset West, South Africa\\
$^{70}$ Jeonbuk National University, Jeonju, Republic of Korea\\
$^{71}$ Johann-Wolfgang-Goethe Universit\"{a}t Frankfurt Institut f\"{u}r Informatik, Fachbereich Informatik und Mathematik, Frankfurt, Germany\\
$^{72}$ Korea Institute of Science and Technology Information, Daejeon, Republic of Korea\\
$^{73}$ KTO Karatay University, Konya, Turkey\\
$^{74}$ Laboratoire de Physique Subatomique et de Cosmologie, Universit\'{e} Grenoble-Alpes, CNRS-IN2P3, Grenoble, France\\
$^{75}$ Lawrence Berkeley National Laboratory, Berkeley, California, United States\\
$^{76}$ Lund University Department of Physics, Division of Particle Physics, Lund, Sweden\\
$^{77}$ Nagasaki Institute of Applied Science, Nagasaki, Japan\\
$^{78}$ Nara Women{'}s University (NWU), Nara, Japan\\
$^{79}$ National and Kapodistrian University of Athens, School of Science, Department of Physics , Athens, Greece\\
$^{80}$ National Centre for Nuclear Research, Warsaw, Poland\\
$^{81}$ National Institute of Science Education and Research, Homi Bhabha National Institute, Jatni, India\\
$^{82}$ National Nuclear Research Center, Baku, Azerbaijan\\
$^{83}$ National Research and Innovation Agency - BRIN, Jakarta, Indonesia\\
$^{84}$ Niels Bohr Institute, University of Copenhagen, Copenhagen, Denmark\\
$^{85}$ Nikhef, National institute for subatomic physics, Amsterdam, Netherlands\\
$^{86}$ Nuclear Physics Group, STFC Daresbury Laboratory, Daresbury, United Kingdom\\
$^{87}$ Nuclear Physics Institute of the Czech Academy of Sciences, Husinec-\v{R}e\v{z}, Czech Republic\\
$^{88}$ Oak Ridge National Laboratory, Oak Ridge, Tennessee, United States\\
$^{89}$ Ohio State University, Columbus, Ohio, United States\\
$^{90}$ Physics department, Faculty of science, University of Zagreb, Zagreb, Croatia\\
$^{91}$ Physics Department, Panjab University, Chandigarh, India\\
$^{92}$ Physics Department, University of Jammu, Jammu, India\\
$^{93}$ Physics Program and International Institute for Sustainability with Knotted Chiral Meta Matter (SKCM2), Hiroshima University, Hiroshima, Japan\\
$^{94}$ Physikalisches Institut, Eberhard-Karls-Universit\"{a}t T\"{u}bingen, T\"{u}bingen, Germany\\
$^{95}$ Physikalisches Institut, Ruprecht-Karls-Universit\"{a}t Heidelberg, Heidelberg, Germany\\
$^{96}$ Physik Department, Technische Universit\"{a}t M\"{u}nchen, Munich, Germany\\
$^{97}$ Politecnico di Bari and Sezione INFN, Bari, Italy\\
$^{98}$ Research Division and ExtreMe Matter Institute EMMI, GSI Helmholtzzentrum f\"ur Schwerionenforschung GmbH, Darmstadt, Germany\\
$^{99}$ Saga University, Saga, Japan\\
$^{100}$ Saha Institute of Nuclear Physics, Homi Bhabha National Institute, Kolkata, India\\
$^{101}$ School of Physics and Astronomy, University of Birmingham, Birmingham, United Kingdom\\
$^{102}$ Secci\'{o}n F\'{\i}sica, Departamento de Ciencias, Pontificia Universidad Cat\'{o}lica del Per\'{u}, Lima, Peru\\
$^{103}$ Stefan Meyer Institut f\"{u}r Subatomare Physik (SMI), Vienna, Austria\\
$^{104}$ SUBATECH, IMT Atlantique, Nantes Universit\'{e}, CNRS-IN2P3, Nantes, France\\
$^{105}$ Sungkyunkwan University, Suwon City, Republic of Korea\\
$^{106}$ Suranaree University of Technology, Nakhon Ratchasima, Thailand\\
$^{107}$ Technical University of Ko\v{s}ice, Ko\v{s}ice, Slovak Republic\\
$^{108}$ The Henryk Niewodniczanski Institute of Nuclear Physics, Polish Academy of Sciences, Cracow, Poland\\
$^{109}$ The University of Texas at Austin, Austin, Texas, United States\\
$^{110}$ Universidad Aut\'{o}noma de Sinaloa, Culiac\'{a}n, Mexico\\
$^{111}$ Universidade de S\~{a}o Paulo (USP), S\~{a}o Paulo, Brazil\\
$^{112}$ Universidade Estadual de Campinas (UNICAMP), Campinas, Brazil\\
$^{113}$ Universidade Federal do ABC, Santo Andre, Brazil\\
$^{114}$ Universitatea Nationala de Stiinta si Tehnologie Politehnica Bucuresti, Bucharest, Romania\\
$^{115}$ University of Cape Town, Cape Town, South Africa\\
$^{116}$ University of Derby, Derby, United Kingdom\\
$^{117}$ University of Houston, Houston, Texas, United States\\
$^{118}$ University of Jyv\"{a}skyl\"{a}, Jyv\"{a}skyl\"{a}, Finland\\
$^{119}$ University of Kansas, Lawrence, Kansas, United States\\
$^{120}$ University of Liverpool, Liverpool, United Kingdom\\
$^{121}$ University of Science and Technology of China, Hefei, China\\
$^{122}$ University of South-Eastern Norway, Kongsberg, Norway\\
$^{123}$ University of Tennessee, Knoxville, Tennessee, United States\\
$^{124}$ University of the Witwatersrand, Johannesburg, South Africa\\
$^{125}$ University of Tokyo, Tokyo, Japan\\
$^{126}$ University of Tsukuba, Tsukuba, Japan\\
$^{127}$ Universit\"{a}t M\"{u}nster, Institut f\"{u}r Kernphysik, M\"{u}nster, Germany\\
$^{128}$ Universit\'{e} Clermont Auvergne, CNRS/IN2P3, LPC, Clermont-Ferrand, France\\
$^{129}$ Universit\'{e} de Lyon, CNRS/IN2P3, Institut de Physique des 2 Infinis de Lyon, Lyon, France\\
$^{130}$ Universit\'{e} de Strasbourg, CNRS, IPHC UMR 7178, F-67000 Strasbourg, France, Strasbourg, France\\
$^{131}$ Universit\'{e} Paris-Saclay, Centre d'Etudes de Saclay (CEA), IRFU, D\'{e}partment de Physique Nucl\'{e}aire (DPhN), Saclay, France\\
$^{132}$ Universit\'{e}  Paris-Saclay, CNRS/IN2P3, IJCLab, Orsay, France\\
$^{133}$ Universit\`{a} degli Studi di Foggia, Foggia, Italy\\
$^{134}$ Universit\`{a} del Piemonte Orientale, Vercelli, Italy\\
$^{135}$ Universit\`{a} di Brescia, Brescia, Italy\\
$^{136}$ Variable Energy Cyclotron Centre, Homi Bhabha National Institute, Kolkata, India\\
$^{137}$ Warsaw University of Technology, Warsaw, Poland\\
$^{138}$ Wayne State University, Detroit, Michigan, United States\\
$^{139}$ Yale University, New Haven, Connecticut, United States\\
$^{140}$ Yonsei University, Seoul, Republic of Korea\\
$^{141}$  Zentrum  f\"{u}r Technologie und Transfer (ZTT), Worms, Germany\\
$^{142}$ Affiliated with an institute covered by a cooperation agreement with CERN\\
$^{143}$ Affiliated with an international laboratory covered by a cooperation agreement with CERN.\\

\end{flushleft} 

\end{document}